\documentclass[conference,peer-reviewed]{conf}

\graphicspath{{./}{figures/}}

\usepackage[utf8]{inputenc}

\usepackage{microtype}

\usepackage[numbers,square]{natbib}

\usepackage{color}
\usepackage{url}

\usepackage{booktabs}

\usepackage{pgfplots}

\usepackage{mathtools}
\usepackage{amsmath}
\usepackage{amssymb}

\DeclareMathAlphabet\mathbfcal{OMS}{cmsy}{b}{n}
\DeclareMathOperator{\median}{median}

\renewenvironment{thebibliography}[1]{%
  \begin{oldthebibliography}{#1}%
    \setlength{\parskip}{0pt}%
    \setlength{\itemsep}{2pt}%
  }%
  {\end{oldthebibliography}}

\title{On the Importance of Temporal Context in Proximity Kernels: A Vocal Separation Case Study}

\author[1]{Delia Fano Yela}
\author[1]{Sebastian Ewert}
\author[2]{Derry FitzGerald}
\author[1]{Mark Sandler}

\affil[1]{Queen Mary University of London}
\affil[2]{Cork School of Music, Cork Institute of Technology}

\correspondence{Delia Fano Yela}{d.fanoyela@qmul.ac.uk}

\lastnames{Fano Yela, Ewert, and Sandler}

\shorttitle{On the Importance of Temporal Context in Proximity Kernels}

\begin{document}

\twocolumn[
\maketitle 

\begin{onecolabstract}

Musical source separation methods exploit source-specific spectral characteristics to facilitate the decomposition process. Kernel Additive Modelling (KAM) models a source applying robust statistics to time-frequency bins as specified by a source-specific kernel, a function defining similarity between bins. Kernels in existing approaches are typically defined using metrics between single time frames. In the presence of noise and other sound sources information from a single-frame, however, turns out to be unreliable and often incorrect frames are selected as similar. In this paper, we incorporate a temporal context into the kernel to provide additional information stabilizing the similarity search. Evaluated in the context of vocal separation, our simple extension led to a considerable improvement in separation quality compared to previous kernels.

\end{onecolabstract}
]

\section{Introduction}

Music recordings typically comprise a mixture of different sound sources, corresponding to musical instruments such as a guitar, drums or vocals. Many applications including up-mixing, automatic transcription or musical feature extraction require or benefit from sources being isolated or enhanced from the rest of the mixture. Even if the signal only contains a single sound source, one may be interested in separating different aspects of the source for further analysis, for example, to differentiate transients related to onsets from the pitched signal components. In this context, various models and techniques have been proposed. Depending on the use case, a major goal is to find characteristics helping with the definition, identification and separation of individual sources. Such characteristics can include various acoustical or perceptual aspects, including the typical behaviour of a source in time (e.g.: vibrato \cite{DriedgerBEM16_VibratoDetection_ISMIR, DriedgerM15_SingingVoice_ICASSP}, continuity in activity \cite{Virtanen07_MonauralSoundSourceSeparation_TASLP, BertinBV10_EnforcingHarmonicityInBayesNMF_TASLP}, repetitiveness of patterns \cite{FitzGerald12_MedianVocal_ISSC, RafiiPardo13_REPET_IEEE-TASLP}) or spectral characteristics (e.g.: broadband vs harmonic energy distribution \cite{Fitzgerald10_HarmPercSep_DAFX} ). These properties are often either modelled explicitly \cite{OzerovVB12_PriorInfoSourceSep_TASLP} or are learned from data \cite{nugraha016_multichannel_TASLP}.

A particularly successful family of techniques is based on Non-Negative Matrix Factorization (NMF) \cite{LeeS00_AlgorithmsNmf_NIPS}, where a time-frequency representation of a signal is modelled as a product of two matrices. The first matrix captures the spectral properties of the signal in its columns, each often referred to as a spectral \emph{template}. The corresponding rows (or \emph{activations}) in the second matrix determine when and how strong each template is present in the signal. As applying the original NMF approach \cite{LeeS00_AlgorithmsNmf_NIPS} to musical data often does not yield useful results \cite{FitzGeraldCC08_ExtendedNMFVariants_CIN}, most of the state-of-the-art source separation methods are based on NMF variants \cite{CichockiZP09_AlternateAlgorithmsNmf_Book} incorporating additional information about the source in the form of spectral constraints \cite{Virtanen07_MonauralSoundSourceSeparation_TASLP, BertinBV10_EnforcingHarmonicityInBayesNMF_TASLP}, user-assisted annotations \cite{Smaragdis09_UserGuidedAudioSelection_ACM-UIST} or score information \cite{EwertPMP14_ScoreInformedSourceSep_IEEE-SPM}. Compared to previous approaches such as Independent Subspace Analysis, NMF relies slightly less on the assumption of statistical independence among sources \cite{AbdallahP04_TranscriptionViaNMF_ISMIR}. However, as a severe limitation in NMF, the results obtained typically strongly depend on how well the spectral templates reflect the actual properties in a given recording. In particular, changes with respect to the instrument or recording conditions can lead to a drastic decrease in separation performance.
 
The recently proposed Kernel Additive Modelling (KAM) \cite{LiutkusFRPD14_KernelAdditive_IEEE-TSP} takes a different approach. Assuming that several sources overlap in a specific bin in a time-frequency representation, the idea is to reconstruct the magnitude for a given source in that bin by analysing the values in other bins, in which the source is likely to assume similar values. The similarity relation between bins is specified by a source-specific kernel, which defines for each pair of bins whether they are to be called similar or not -- typically using some sort of underlying metric in the background. The goal in KAM is to design a source-specific proximity kernel that indicates, given a specific bin, where to look for similar bins in a time-frequency representation of the signal. Once a set of similar bins has been identified for a source, the contribution of the remaining sources can be regarded as outliers -- assuming that not all sources have the exact same kernel function. This way, KAM produces an estimate of the magnitude in a bin for a specific source by applying robust statistics (typically the median) across the similar bins.

For vocal separation, existing KAM instances \cite{FitzGerald12_MedianVocal_ISSC,RafiiPardo13_REPET_IEEE-TASLP} considered the accompaniment to be far more stationary and repetitive than the vocal source. This means that there are many time frames containing the same (or similar) accompaniment but not many with the same voice content. Further, it is implicitly assumed that the energy contribution of the accompaniment is higher than that of the vocal source. In line with this reasoning, the kernel proposed in \cite{FitzGerald12_MedianVocal_ISSC} is a function that, based on the Euclidean distance, returns the \emph{K} frames most similar to a given frame. More precisely, in this case, a bin is considered similar to a second bin if both have the same centre frequency and the frame number for the second is among the K most similar frames. Based on the above assumptions, the voice thus can be regarded as an outlier and can be eliminated using median filtering across the similar bins.

Even though this kernel can exploit some of the source's regularities, its simplicity leads to some drawbacks. To illustrate this, let's consider a recording containing two instrumental solo sections for a guitar and a piano. Depending on the recording conditions, the sustain part for both instruments can have a similar energy distribution in frequency direction (playing the same musical pitch). As a consequence, a frame-wise kernel based on the Euclidean distance sometimes fails to identify the actual dissimilarity between frames and can confuse a guitar frame with a piano frame. Such issues are even more pronounced if an instrument has variable timbre, for example due to the use of effects. This mix-up can lead to an unexpected energy distribution for an instrument in the separation result. 

Using only a single frame, such issues are difficult to resolve. However, by taking the temporal context of a frame into account, we obtain more information about which frames are actually similar to each other. For example, using a larger temporal context, the similarity measure might take a frame containing the onset into account, which can be very discriminative for an instrument. Also, the temporal context might even be large enough to pick up some basic information about the musical context and, assuming the different instruments play different note patterns, we can use this low-level musical context as additional guidance to find similar frames for a given instrument. 

Based on this simple idea, we propose in this paper to modify existing kernels by introducing a temporal context. Basically, given a frame we aim to find similar frames for, we include the preceding and succeeding frames in the similarity function underlying our kernel. Effectively, that means we measure similarity based on entire groups of frames instead of single frames. The size of the temporal context is chosen large enough to give some rough indication of local musical patterns. Re-using the previous guitar-piano example and assuming that the current frame is in the guitar solo, the group of frames centered around this frame might span a few notes. Now, when looking for similar segments, we can take this local note constellation to some degree into account, which potentially aids in differentiating between similar timbres. In particular, unless the solo piano section contains the same sequence of notes played in the same fashion, the guitar will not be mistaken for the piano. 

The remainder of the paper is organized as follows. In Section~\ref{sec:relatedWork} we give a brief overview of related work, followed by Section~\ref{sec:method}, where we describe the details of our proposed extension. Next, in Section~\ref{sec:experiments} we report on experiments indicating the level of improvement resulting from our extension. Finally, we conclude in Section~\ref{sec:conclusion} with an outlook to possible research directions.

\section{Related Work}
\label{sec:relatedWork}

Methods for musical source separation typically incorporate various types of prior knowledge about the individual sound sources or the mixing process. In a vocal separation scenario, one can exploit various properties of the singing voice and of the background or accompaniment. Some methods start by differentiating between vocal and non-vocal regions. For example, the method presented in \cite{raj07_SingerBackground_FRSM} uses a priori knowledge about non-vocal segments to learn an accompaniment model based on Probabilistic Latent Component Analysis (PLCA) and then fixes the accompaniment to learn the vocal source. The methods presented in \cite{ozerov07_BayesianVoiceMusic_TASLP,li07_VoxAccompSep_TASLP} employ Mel Frequency Cepstral Coefficients (MFCC) and Gaussian Mixture Models (GMM) to first differentiate between vocal and non-vocal regions and then use the resulting information to train a Bayesian model for the accompaniment \cite{ozerov07_BayesianVoiceMusic_TASLP}. Using similar pre-processing steps, the approach introduced in \cite{li07_VoxAccompSep_TASLP} extracts the vocal pitch contour using a predominant pitch estimator on the vocal segments and performs separation through binary masking. Similarly, \cite{hsu10_improvementVoxSep_TASLP} uses a predominant pitch estimator based on a Hidden Markov Model (HMM) to extract the vocal pitch contour through spectral subtraction in voiced segments and GMMs to identify and separate the unvoiced consonants. 

Another large body of work is based on Deep Neural Networks (DNNs) that are typically trained on the magnitude spectrogram of the mixture to predict either a time-frequency mask describing the energy distribution of a source relative to the other sources \cite{narayanan13_idealMaskDNNSpeech_ICASSP} or the source spectrogram directly \cite{huang14_singingDNN_ISMIR,nugraha016_multichannel_TASLP}. In \cite{nugraha016_multichannel_TASLP} the authors employ a DNN to extract a spectrogram for each source using multi-channel recordings as input; the parameters of a multi-channel Wiener filter are estimated using an iterative Expectation-Maximization (EM) algorithm. While most state-of-the-art methods for vocal separation employ some variant of DNNs, these methods are typically trained for specific combinations of instruments or instrument groups, which limits their flexibility and adaptivity in practice. Further, the performance of these techniques depends strongly on the quality of the training data.

Instead of training a model, various methods target the inherent properties of the sources directly and use their differences to distinguish them in the separation process. For instance, the repetition of patterns is a core characteristic of popular music. In this context, some methods \cite{huang12_singingRPCA_ICASSP,chan15_vocalActivityInformed_ICASSP} employ Robust Principal Component Analysis (RPCA) to decompose the spectrogram into a low-rank and a sparse matrix, and argue that these can be associated with the accompaniment and vocals, respectively. The method presented in \cite{chan15_vocalActivityInformed_ICASSP} takes this idea a step further by introducing vocal activity information into the standard RPCA algorithm.

For cases in which the background music can be considered to be repetitive, the method REPET \cite{RafiiPardo13_REPET_IEEE-TASLP} identifies the repetition period of the musical pattern, models the repeating segment and, through the use of robust statistics and soft-masking, extracts the repeating pattern associated with the background music. Even though it has proven successful in a variety of contexts, it is limited to only one repeating pattern and is unable to further differentiate individual sound sources. The method \cite{RafiiPardo13_REPET_IEEE-TASLP}, as well as others relying on a similarity measure such as \cite{FitzGerald12_MedianVocal_ISSC, FanoYelaEFS17_HybridKamNmf_ICASSP}, can be considered as instances of the more general KAM framework \cite{LiutkusFRPD14_KernelAdditive_IEEE-TSP} described in the introduction. The approach presented in this paper is also a member of the KAM family of methods as it represents an extension to the method described in \cite{FitzGerald12_MedianVocal_ISSC}.

\section{Proposed Method}
\label{sec:method}

\begin{figure*}[ht]
\centering
\includegraphics[width=\textwidth]{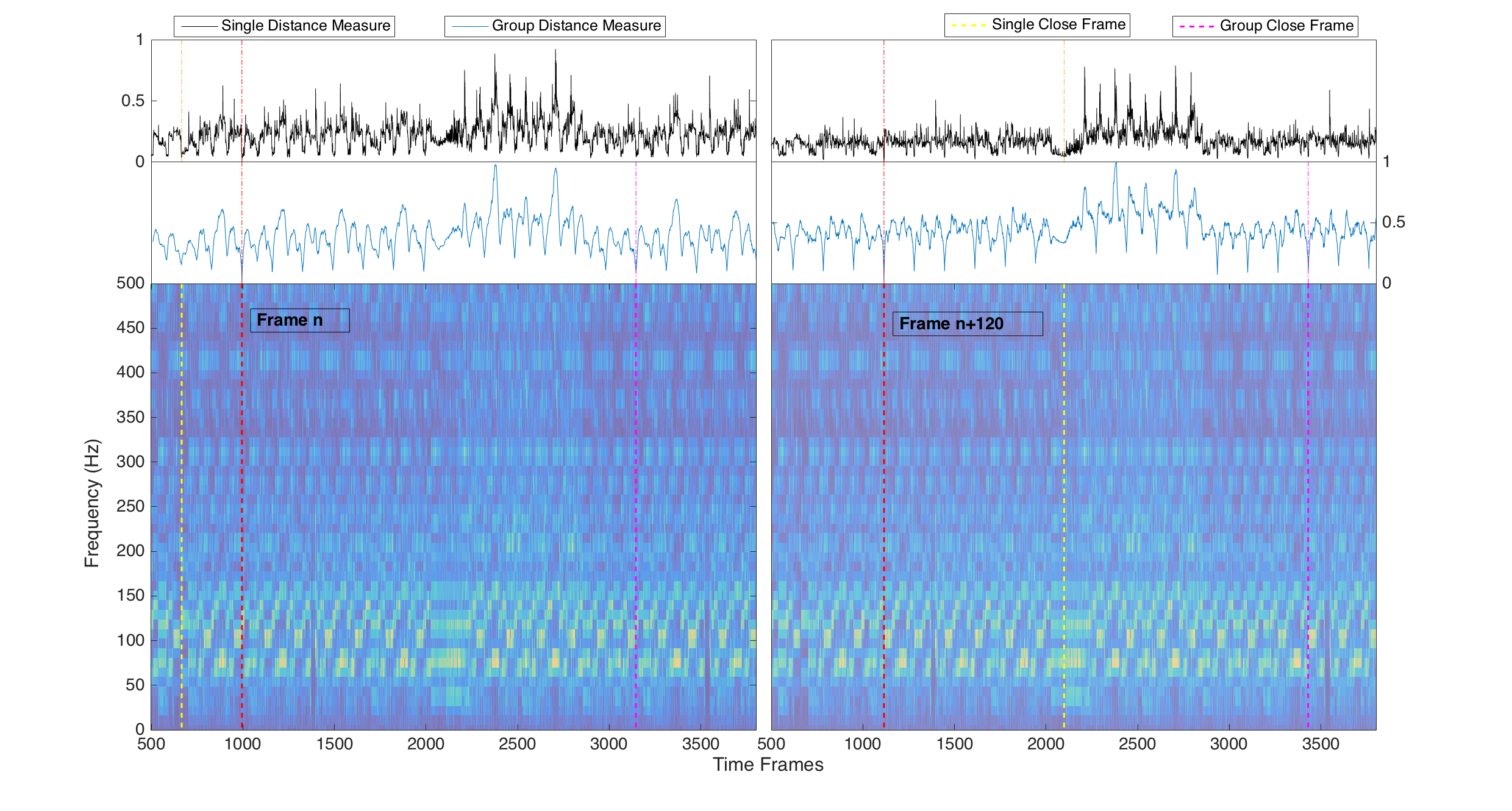}
\vspace{-0.9cm}
\caption{Comparison between single and group distance measure: given a spectrogram of the mixture (shown in each half of the figure), the plots show the distance between a specific frame ($k = 1000$ on the left and $k=1120$ on the right, see red vertical lines) and the remaining frames. The single-frame and group distance measure are plotted in black and blue, respectively. The frames closest with respect to these distances are indicated using yellow (single-frame) and magenta (group) vertical lines.}
\label{fig:distance}
\end{figure*}

To describe our approach, we follow the notation used in \cite{FitzGerald12_MedianVocal_ISSC}. The method can be regarded as an instance of KAM using only one iteration of the Kernel Backfitting procedure described in \cite{LiutkusFRPD14_KernelAdditive_IEEE-TSP}.

\begin{figure*}[ht]
\centering
\includegraphics[width=2\columnwidth,height=\columnwidth]{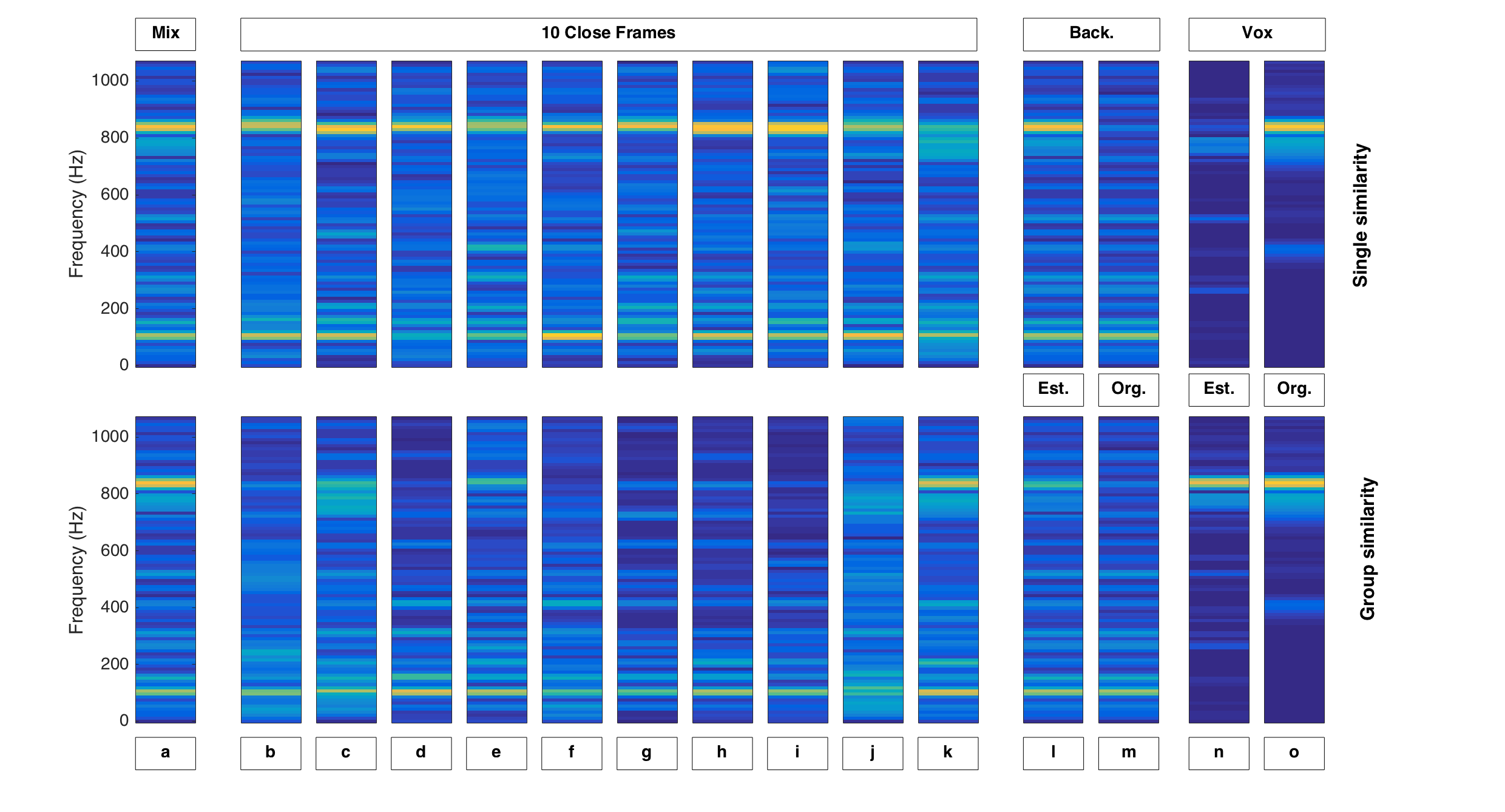}
\vspace{-0.5cm}
\caption{ Given a mixture frame (a), the first ten close frames (b)-(k) identified using single and group similarity yield a background music (l) and vocal (n) approximations of the clean sources, (m) and (o) respectively. }
\label{fig:median}
\end{figure*}

More precisely, let $C \in \mathbb{C}^{M \times N}$ be the spectrogram of a music recording containing a mixture of a vocal and an accompaniment track, where $M$ is the number of frequency bins and $N$ the number of time frames, see Fig.~\ref{fig:distance}. Further, let $X \in \mathbb{R}^{M \times N}$ be the corresponding magnitude spectrogram. As a first step, the vocal separation algorithm proposed in \cite{FitzGerald12_MedianVocal_ISSC} computes for each pair of frames $(k,\ell) \in \{1,\ldots,N\}\times \{1,\ldots,N\}$ the Euclidean distance between the two corresponding columns in $X$:
\[
D_{k,  \, \ell}  = \displaystyle\sum_{m \, =  \, 1}^{M} ( X_{m,  \, k} - X_{m, \, \ell} ) ^{2}.
\]
In Fig.~\ref{fig:distance} the $k$-th row of $D$ is plotted in black, with $k = 1000$ in the left and $k=1120$ in the right half of the figure. The spectrogram of the mixture to be processed is shown in each half of the figure and the two frames are indicated by a vertical red line.

To obtain a list of $P$ frames being closest to a given frame $k$, the symmetric matrix $D$ is first sorted individually in each row. By keeping track of which frame index belongs to which entry in the sorted matrix $D$, we can create for each $k$ a matrix $A^k \in \mathbb{R}^{M \times P}$ in which the $P$ columns contain the $P$ closest frames, i.e. a specific subset of frames taken from $X$. This process is illustrated in the left part of Fig.~\ref{fig:distance}, where the most similar frame (indicated by a yellow vertical line) is found using the similarity values in $D$. For the upper part of Fig.~\ref{fig:median}, this process was repeated until for a given frame (shown in Fig.~\ref{fig:median}a) the $P=10$ closest frames were found (shown in Fig.~\ref{fig:median}b-k), i.e. these 10 frames represent the first ten columns of $A^k$.  

Assuming that the energy in each frame and hence the Euclidean similarity measure are dominated by the accompaniment, we now have $P$ frames similar to frame $k$ that only differ in terms of the vocal part (following above assumptions). In particular, the vocal part leads to outliers and we want to extract the commonalities between the frames in $A^k$. To this end, the method in \cite{FitzGerald12_MedianVocal_ISSC} employs the median filter, which is invariant against outliers (up to 50 percent) and belongs to the class of operator used in robust statistics. More precisely, we define the estimated magnitude spectrogram $Y \in \mathbb{R}^{M \times N}$ as follows:
\[
Y_{m, \, k} := \median( A^k_{m,1},\ldots,A^k_{m,P} )
\]
To extract both magnitude and vocals from the mixture, we create a mask (similar to a Wiener filter) as in \cite{FitzGerald12_MedianVocal_ISSC}. More precisely, we measure the distance between the mixture $X$ and the accompaniment estimate $Y$ after a logarithmic compression (with the logarithm leading to a perceptually more meaningful distance \cite{FitzGerald12_MedianVocal_ISSC}) and employ this distance in a Gaussian radial basis function to obtain a mask $W \in [0,1]^{M\times N}$:
\[
W_{m,n} = \exp\left(-\frac{(\log(X_{m,n}) - \log(Y_{m,n}))^{2}}{2\lambda^{2}}\right),
\]
where $\lambda$ is a parameter to additionally compress the log-distances non-linearly. In the following, we set $\lambda$=1.
The complex spectrograms for the accompaniment $B \in \mathbb{C}^{M \times N}$ and vocals $V \in \mathbb{C}^{M \times N}$ can then be estimated by applying the soft masks $W$ and $(1-W)$ to the original mixture spectrogram $C$ using an element-wise multiplication $\odot$, respectively:
\[
B = W \odot C 
\] 
\[
V = (1 - W) \odot C
\]
With this framework in place, our extension can be explained in a simple, straightforward way. In particular, we keep the same basic procedure but replace $D$ with a new pairwise distance matrix $\widetilde{D}$, taking a temporal context for each pair of frames additionally into account. More precisely, we define:
\[
\widetilde{D}_{k,  \, \ell}  = \displaystyle\sum_{m \, =  \, 1}^{M}\sum_{r \, =  \, -R}^{R} ( X_{m,  \, k+r} - X_{m, \, \ell+r} ) ^{2},
\]
where $R\in\mathbb{N}$ is a radius in number of frames defining the extend of our temporal context. This way, we do not compare single frames anymore but whole groups of frames (resembling concepts used in the context of musical structure analysis \cite{MuellerK06_EnhancingSimilarityMatrices_ICASSP}).

Overall, this change in the distance matrix is a rather small extension -- however, we found this small change to have a reasonably strong impact on the separation result. To illustrate our findings, we discuss Figures~\ref{fig:distance}~and~\ref{fig:median} now in more detail. We start with the first row in Fig.~\ref{fig:median}. Fig.~\ref{fig:median}a shows a frame taken from the input $X$ that we wish to process. As we can see, the frame contains two strong partials, one corresponding to the background music (100Hz, compare also Fig.~\ref{fig:median}m for the ground truth) and another one related to the vocal source (830Hz, compare Fig.~\ref{fig:median}o). Using the single-frame distance $D$, we obtain results equivalent to the KAM-based baseline \cite{FitzGerald12_MedianVocal_ISSC}: the 10 closest frames, taken from $A^k$, are shown in Figs.~\ref{fig:median}b-k.
As we can see, due to the strong vocal activity in the input, the single frame distance used in $D$ is heavily influenced by the vocal partials and frames are selected in Figs.~\ref{fig:median}b-k that are also dominated by similar vocal activity. As a consequence, even applying the median filter to these frames in $A^k$ does not help with the identification of the vocal partial as an outlier -- simply because the vocal partial occurs in every frame. Comparing the median filtered result for the accompaniment (Fig.~\ref{fig:median}$\ell$, computed using $P=100$) with the ground truth (Fig.~\ref{fig:median}m), we observe that the vocal partial remains intact and the separation was ineffective -- compare also Fig.~\ref{fig:median}n, which contains the vocal estimate which hardly contains energy. We have identified this problem to appear consistently in scenarios with a low Signal-to-Noise Ratio (SNR), taking the signal as the source we wish to isolate and the remaining sound sources as noise. This behavior represents a considerable drawback in the current kernel design proposed in KAM. 

Once we introduce a temporal context in the distance function, the importance of the frame to be filtered is lowered while the importance of having a similar neighbourhood is increased.
To see this, we now look at the second row in Fig.~\ref{fig:median}, where the results are shown using our extension. Taking a look at the 10 closest frames (Figs.~\ref{fig:median}b-k), we notice a higher diversity among them compared to the previous scenario -- however, the vocals are a lot less dominant while the accompaniment is more prominent. Looking at the result after the median filter (Fig.~\ref{fig:median}$\ell$, computed using $P=100$), we see that there is still some vocal energy left but, compared to the single-frame distance, the vocals are much more suppressed and the result is considerably closer to the ground truth (Fig.~\ref{fig:median}m). The improvement is also clearly visible in the separation result for the vocals (compare Fig.~\ref{fig:median}n and o).

It is also interesting to compare the two distances directly in the form of the matrices $D$ and $\widetilde{D}$. Fig.~\ref{fig:distance} shows the row for frame $k$ in $D$ (plotted in black) and in $\widetilde{D}$ (plotted in blue), with $k=$1000 in the left part of the figure and $k=$1120 for the right part. As we can see, the single frame distance is much more noisy compared to the one with temporal context. Also, peaks indicating a low distance (i.e. high similarity) are much clearer for the curve using a temporal context -- this is particularly visible in the right half of the figure where many spurious peaks can be found in the single frame distance. This overall change in qualitative behaviour also influences which frames are selected as the most similar frames. In particular, the yellow and magenta lines in Fig.~\ref{fig:distance} indicate the most similar frame found using the single frame and group of frames distance, respectively.

Comparing the yellow and magenta position in the spectrogram on the left, we see that both indicate a frame with a low distance that additionally makes sense musically as both happen at the end of a similar note constellation.
In the right half of the figure ($k=$1120), however, we can notice that the frame selected via the single frame distance is in a completely different section of the song. Zooming in, we find that the magnitude values are indeed similar which explains this selection but, being in a different section of the song, there are various subtle differences in that frame leading to additional difficulties for the median filter. Looking at the distance values using the temporal context, we see that around the hit for the single-frame distance, the distance values are here quite high, which indicates a musical dissimilarity to the pattern around location $k=$1120.

\section{Experiments}  
\label{sec:experiments}

To quantitatively compare our proposed extension with the baseline \cite{FitzGerald12_MedianVocal_ISSC}, we employed the Demixing Secrets Dataset 100 (DSD100) as also used in the 2016 Signal Separation Evaluation Campaign (SiSEC) \cite{liutkus20172016}. The dataset contains 100 different songs of various genres, all of them being polyphonic, mixed in stereo, with a sampling frequency of 44.1 kHz and 30s long. In order to asses the separation quality, we used the toolkit available in the SiSEC website, which employs the BSS Eval toolbox 3.0 \cite{VincentGF06_PerformanceMeasurement_IEEE-TASLP} to calculate the Signal to Distortion Ratio (SDR), Source Image to Spatial Distortion Ratio (ISR), Source to Interference Ratio (SIR) and the Source to Artifacts Ratio (SAR). 

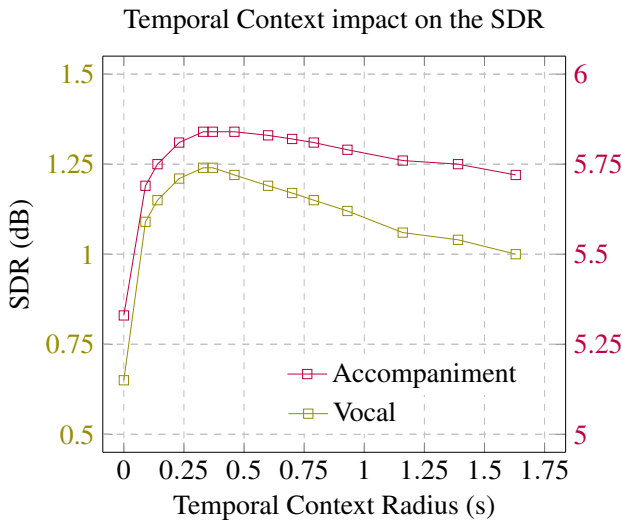
\begin{figure}[t]
\centering
\scalebox{0.87}{
	\pgfplotsset{every axis/.append style={font=\large}}
	\begin{tikzpicture}
	\hspace{-0.8cm}
		\begin{axis}[ 
			width=7cm,
    			title={Temporal Context impact on the SDR},
    			scale only axis,
			xmin=0, xmax=1.75,
    			ymin=0.5, ymax=1.5,
			enlargelimits=0.05,
			xtick={0, 0.25, 0.5, 0.75, 1, 1.25, 1.5, 1.75},
    			ytick={0.5, 0.75, 1, 1.25, 1.5},
                		yticklabels = {\textcolor{olive}{0.5}, \textcolor{olive}{0.75}, \textcolor{olive}{1}, \textcolor{olive}{1.25}, \textcolor{olive}{1.5}},
			axis y line*=left,
    			xlabel={Temporal Context Radius (s) },
    			ylabel={SDR (dB)},
                		axis y label/.style= {at={(ticklabel cs:0.5)},rotate=90,anchor=near ticklabel, yshift=-9pt},
			legend style={draw=none, , at={(0.66, 0.15)}},
    			xmajorgrids=true,
			ymajorgrids=true,
    			grid style=dashed] 
			\addplot[ color=olive, mark=square]
    			coordinates {(0, 0.65)(0.09, 1.09)(0.14, 1.15)(0.23, 1.21)(0.33, 1.24)(0.37, 1.24)(0.46, 1.22)( 0.6, 1.19)( 0.7, 1.17)( 0.79, 1.15)( 0.93, 1.12)					( 1.16,1.06)( 1.39, 1.04)(1.63, 1.00) };
   			 \legend{Vocal}
		\end{axis}
		\begin{axis}[
    			width=7cm,
			scale only axis,
    			xmin=0, xmax=1.75,
    			ymin=5, ymax=6,
			enlargelimits=0.05,
    			ytick={5, 5.25, 5.5, 5.75, 6},
                		yticklabels = {\textcolor{purple}{5}, \textcolor{purple}{5.25}, \textcolor{purple}{5.5}, \textcolor{purple}{5.75}, \textcolor{purple}{6}},
			axis y line*=right,
            		axis x line=none,
    			legend style={draw=none, at={(0.92, 0.25)}},
    			ymajorgrids=true,
    			grid style=dashed]
			\addplot[color=purple, mark=square]
    				coordinates {
    				(0, 5.33)(0.09, 5.69)(0.14, 5.75)(0.23, 5.81)(0.33, 5.84)(0.37, 5.84)(0.46, 5.84)( 0.6, 5.83)( 0.7, 5.82)	( 0.79, 5.81)( 0.93, 5.79)( 1.16, 						5.76)( 1.39, 5.75)(1.63, 5.72)};
    			\legend{Accompaniment}
		\end{axis}
	\end{tikzpicture}}
\caption{SDR results for the proposed extension with different temporal contexts for the DSD100 dataset. }
\label{fig:radiusplot}
\end{figure}

For this evaluation, we have implemented (using an FFT size of 4096 and a hopsize of 2048 samples) an instance of KAM for vocal separation following \cite{FitzGerald12_MedianVocal_ISSC} and introduced a temporal context in the proximity kernel as described in Section~\ref{sec:method}. The number of frames $R$ specifying the temporal context is a parameter of our approach.
In principle, this parameter could be adapted for every frame based on musical knowledge, for example, based on segmentation information or pitch tracking data, which would render the method more flexible and adjustable to musical changes in the signal. However, for this paper, we chose to use a fixed setting for the radius $R$. To find a suitable value, we conducted a simple parameter sweep, whose results are shown in Fig.~\ref{fig:radiusplot}. The figure shows the averaged SDR values for both vocal and accompaniment separation using the proposed extension for different radius values. We can observe an overall trend in Fig.~\ref{fig:radiusplot} shared by both vocal and accompaniment separation, where the biggest difference in SDR value is between a zero radius (the baseline method) and the other values taking a temporal context into account. In addition, we see that the highest SDR values are achieved for a temporal context of around 1 second (radius values between 0.25 and 0.6 seconds), which can be considered wide enough to capture some simple musical patterns. If the radius is increased, the musical information within the temporal context grows and we observe a slight decrease of the SDR. For this reason, we chose to fix the radius to 372ms.

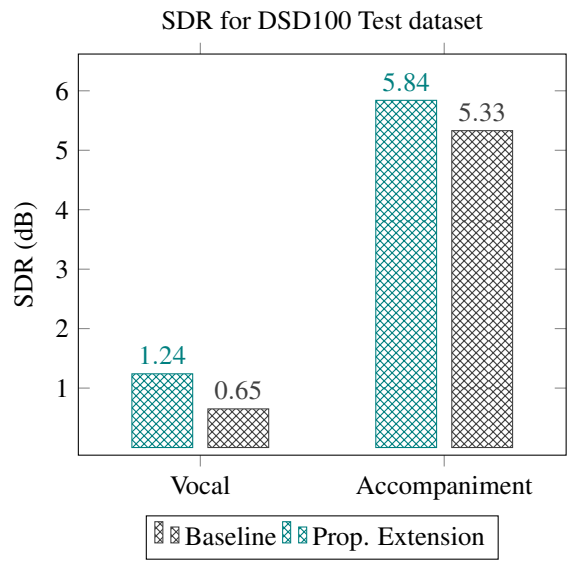
\begin{figure}
	 \usetikzlibrary{patterns}
	\begin{tikzpicture}
		\begin{axis}[
			width=8cm,
			title={SDR for DSD100 Test dataset},
			ybar,
			enlargelimits=0.15,
			legend style={at={(0.5,-0.15)},
			anchor=north,legend columns= -1},
			ylabel={SDR (dB)},
			ytick ={1,2,3,4,5,6},
			every axis y label/.style=
			{at={(ticklabel cs:0.5)},rotate=90,anchor=near ticklabel},
			symbolic x coords={Vocal, Accompaniment},
			xtick= {Vocal, Accompaniment},
			nodes near coords,
			nodes near coords align={vertical},
			enlarge x limits= 0.5,	
			bar width  = 0.8cm,			
			]
			\addplot	+[bar shift= 0.5cm, color = darkgray, pattern=crosshatch, pattern color = darkgray]
			coordinates {(Vocal, 0.65) (Accompaniment, 5.33)};
			\addplot +[bar shift=  -0.5cm, color = teal, pattern=crosshatch, pattern color = teal]
			coordinates {(Vocal, 1.24) (Accompaniment, 5.84)};
			\legend{Baseline, Prop. Extension}
		\end{axis}
	\end{tikzpicture}
\caption{SDR results for the baseline method and the proposed extension tested on the DSD100 dataset.}
\label{fig:sdr_bar}
\end{figure}

Using this fixed value for $R$, Figure~\ref{fig:sdr_bar} shows the SDR values comparing our proposed method to our baseline \cite{FitzGerald12_MedianVocal_ISSC} (i.e. using single frame distances) in more detail. On the SiSEC dataset, our proposed method consistently outperforms the baseline and improves the results by about 0.5dB SDR on average for both vocal and accompaniment separation. Given the simplicity of our extension, this is quite considerable.

Overall, the results are encouraging for such a simple unsupervised method that requires no prior training. Even though there is still room for improvement, introducing temporal context in the similarity search has shown clear advantages.

\section{Conclusion}
\label{sec:conclusion}
We presented a simple approach to improve the similarity search in proximity kernels as used in the KAM framework for source separation. We motivate the need for introducing a temporal context in kernels by analyzing different scenarios where the similarity search would fail otherwise. The results obtained show an improvement in separation performance compared to the baseline on the DSD100 dataset used in SiSEC 2016.
Our results indicate that our extension to the similarity measure temporally stabilises the source estimates and improves the separation performance over the baseline algorithm. Possible future directions include a more extensive study of different approaches to adaptively set the length of the temporal context, taking source-specific characteristics into account on a frame-by-frame level.

\textbf{Acknowledgement:} This work was funded by EPSRC grant EP/L019981/1.

\small
\bibliographystyle{IEEEtranModSEAbbrvIEEE}

\bibliography{referencesMusic}

\end{document}